# AI Thinking:
# A framework for rethinking artificial intelligence in practice

Denis Newman-Griffis
Information School, University of Sheffield, Sheffield, UK
Research on Research Institute, London, UK

**Abstract**

Artificial intelligence is transforming the way we work with information across disciplines and practical contexts. A growing range of disciplines are now involved in studying, developing, and assessing the use of AI in practice, but these disciplines often employ conflicting understandings of what AI is and what is involved in its use. New, interdisciplinary approaches are needed to bridge competing conceptualisations of AI in practice and help shape the future of AI use. I propose a novel conceptual framework called AI Thinking, which models key decisions and considerations involved in AI use across disciplinary perspectives. The AI Thinking model addresses five practice-based competencies involved in applying AI in context: motivating AI use in information processes, formulating AI methods, assessing available tools and technologies, selecting appropriate data, and situating AI in the sociotechnical contexts it is used in. A hypothetical case study is provided to illustrate the application of AI Thinking in practice. This article situates AI Thinking in broader cross-disciplinary discourses of AI, including its connections to ongoing discussions around AI literacy and AI-driven innovation. AI Thinking can help to bridge divides between academic disciplines and diverse contexts of AI use, and to reshape the future of AI in practice.

## 1. Introduction

Artificial intelligence has been positioned as a key driver of global change through a fourth industrial revolution, and AI advances are actively transforming how we process, analyse, and learn from information (1–3). The increasing relevance of AI across disciplines and sectors has led to a wide variety of views on the nature of this transformation, but it is most often envisioned through a technological lens: self-contained AI technologies that change (or replace) the processes we use to work with information. However, the technology-centric perspective that drives much of AI development and innovation fails to capture important aspects of the complex,

*Correspondence to: Denis Newman-Griffis, d.r.newman-griffis@sheffield.ac.uk*

human contexts in which AI is used, and is increasingly insufficient to address the challenges of effectively and ethically using AI in real-world settings (4–6).

The ongoing changes in the AI landscape are not only affecting the scope and scale of AI use: they are also reshaping what it looks like to work with AI technologies in practice. For most of its history, AI has been a specialised, technical discipline, focused on methodological innovation and requiring advanced training and significant computational resources to use effectively. As technologies have matured, AI is in the midst of an expansion from expert technique to everyday toolkit, opening up new opportunities for its use to ask novel questions and perform complex analyses without the need for deep technical expertise or innovation (7,8). This growth in the purposes for which AI is used has outpaced the professional training and development of best practice required to ensure that AI use is both effective and ethical. This has left significant knowledge gaps in the specific skills and competencies involved in using AI in practice, especially in newer and more interdisciplinary contexts, and the specific skills and competencies needed to work with AI as part of processing information.

Part and parcel of these knowledge gaps, and a key challenge created by the growth of AI application and research, is the fact that different disciplines, fields, application contexts, and stakeholders have different conceptualisations of what AI is and what is involved in using it. Each discipline in which AI is studied or used tends to define and operationalise AI in its own ways, with explicitly interdisciplinary approaches to AI in practice being few and far between (9–11). These different conceptualisations of AI in disciplinary literature and applied contexts are rarely surfaced and often in conflict with one another, functioning as competing epistemologies rather than complementary viewpoints. As AI use becomes more urgently interdisciplinary, by virtue of the diversity of AI users and the breadth of impact from AI applications, we need strategies to help partially reconcile these different understandings of AI in practice, and to work effectively across varied perspectives on AI to develop shared practices for AI use.

In this article, I propose a novel framework, which I call *AI Thinking*, which holistically models key decisions and considerations involved in bringing AI into practical use. AI Thinking is a conceptual, competency-based model of the practices involved in applying AI in context, designed to help users of AI systems ensure that their applications are both goal-driven and context-sensitive. To do this, I approach AI as an information *methodology,* rather than an information *technology*, in the process describing and affecting how we work with information as a whole. AI Thinking bridges disciplinary perspectives, illustrating how the design and implementation of AI systems is inextricable from the practical contexts they operate in, and how the social systems surrounding AI are reshaped by the process of computation. By providing practice-based connections between different disciplines and practical contexts, AI Thinking encourages users to work with AI as an interconnected set of technical processes and social practices. By framing AI use in terms of the practices that connect specific problems, technologies, and contexts, AI Thinking highlights shared concerns and reference points that help connect different definitions and practical understandings of AI. The AI Thinking framework thus makes clearer links between academic disciplines and across diverse contexts of AI use, and can help to lower communication costs between different perspectives on what AI means and how to understand, train people in, and manage its use.

## 2. Bridging AI divides across disciplines and contexts

To illustrate the need for a more interdisciplinary conceptualisation of how we work with AI in practice, I first outline key differences in the broad range of conceptualisations of AI. The aim of the AI Thinking model is to bridge between competing (and sometimes incompatible) perspectives and approaches across disciplines, including computer science and engineering, science and technology studies, critical data/AI studies, and the broad landscape of AI applications. Negotiating the varying conceptualisations adopted in these fields, and the different visions they present of what matters most in studying and managing AI, is one of the most significant current challenges in responsibly applying and managing AI in practice. The competency-based model of AI Thinking presented in this article reflects on these varied understandings together with the growing body of evidence on AI practices in diverse settings, including research and innovation, business, health, and policy (12–18). In the sections below, I briefly outline major differences in AI understandings and the historical contexts from which AI practices and the AI Thinking model have emerged.

### 2.1 Operational terminology

The current environment of rapid change in AI technologies and applications is also one of rapid change in language. The range of disciplinary perspectives addressed in this article make it necessary to establish an operational terminology with which to describe AI in practice, and to set the scope of what aspects of AI technologies and practice are in focus.

***AI use:*** I focus in this article on the use of AI for information processing—eg, transforming and analysing data, learning from evidence, and informing action—common across many applications in business, research, policy, healthcare, and other decision-making settings. The AI Thinking framework has potential to inform more creative uses of AI, eg for content generation or creative practice, and may prove helpful in exploring evolving practice in this area. This focus on information processing does not exclude the use of generative AI, which has significant (and developing) applications for both information processing and creative use (cf (19,20)).

***AI technologies:*** Artificial intelligence, as a family of technologies, remains notoriously difficult to define. The evolving scope of AI technologies—ie, specific models, algorithms, and technological products—includes interactive systems such as generative AI, foundation models for adapting general-purpose knowledge, and more specialised AI systems (sometimes referred to as "narrow" AI) targeting specific applications through bespoke machine learning or expert systems (21,22). The use of "AI" in this article does not distinguish between these types of AI technologies, but focuses on common processes and ways of thinking that inform all of these types of AI in practice.

***AI systems:*** AI use is often discussed in terms of "AI systems," but the nature and boundaries of these systems have varied definitions. In many cases, for example in industrial practice (23) or international policy contexts (22), AI systems are defined in terms of software and technical systems alone. In presenting the framework of AI Thinking, I follow thinking in critical data studies by taking a wider view of the contexts in which AI software and technologies are inextricably situated, and recognising the essential role of non-technical stakeholders and perspectives in setting the shape and the purpose of AI technologies (24). Attending to the broader milieu in which AI operates calls for a systems thinking approach (25), and reflects the inherently sociotechnical nature of AI (6). Thus, from an AI Thinking perspective, an 'AI system' includes not only the software and technologies through which AI computation occurs, but also the broader system of people, data, and processes which motivate and make use of AI computation.

### 2.2 Differences in AI understandings across disciplines

The historical contexts of different fields and areas of AI application have naturally led to different conceptualisations of the same ideas. These differences, which are not always immediately visible, hinder the interdisciplinary dialogue and collaboration necessary to make AI use ethical, explainable, and effective in practice (7,26,27). Differences especially in the language around AI and the scope of who is involved in its application can also reflect deeper underlying differences in how AI is conceptualised between disciplines and contexts, and what aspects of AI systems are considered most important.

In my proposed AI Thinking framework, I aim to provide points of common reference and comparison to work across different understandings of AI by joining up technical and social practices involved in AI use. Here, I briefly outline three common ways of framing AI observed across different disciplinary contexts: AI as a linear, cyclical, and relational system. I show how an AI Thinking approach can help to bridge their distinct perspectives on data, computation, and the scope of AI systems.

A ***linear*** understanding of AI, commonly encountered in computer science and engineering disciplines, focuses on the process of computation: beginning with input data, performing a sequence of transformations and analyses, and producing output for further use. On this framing, computation is emphasised and data are pre-existing, material objects to be consumed, transformed, and/or generated. AI systems are typically considered through a technological lens, as separate concerns from the production of data or use of AI outputs. This perspective is reflected in the directional flow diagrams commonly used to illustrate AI model structures and data analysis in computing literature (cf (28,29)) and some applications (30).

A ***cyclical*** understanding of AI, commonly seen in AI applications such as in healthcare or finance, takes a more process-oriented view of AI use, including data collection, analysis, interpretation and assessment, and action taken in response, which frequently produces new data for a subsequent cycle. On this framing, computation is one piece of a larger process which forms the broader context for AI systems as embedded technologies. Data are emphasised, as representative of the information being moved, understood, and acted upon in the overarching process. This perspective is reflected in visualisations and discussion of AI systems in professional practice (31,32) and in the growing literature on human-AI teaming (33,34).

A ***relational*** understanding of AI, more commonly encountered in social sciences and humanities disciplines, focuses on the people, perspectives, and purposes behind the use of AI in sociotechnical contexts. A wide range of theoretical approaches are used to investigate AI from more relational perspectives, including systems thinking (25) and assemblage theory (35), as well as more specific models such as data journeys (36). On this framing, computation is frequently a means rather than a process, emphasising what Amoore describes as 'the capacity to generate an actionable output from a set of attributes' (37), and the data that describe these attributes are a dynamic and contested site of practice. This perspective is frequently reflected in critical studies of AI and analyses of AI harms (38,39). The definition of AI systems adopted for this article most closely aligns with a relational understanding of AI.

To create a cohesive model for AI use across disciplines, these differing understandings of AI must be treated as what they are: equally necessary and informative for a full picture of AI Thinking. The *linear* framing is essential for choosing between and working with specific AI technologies, which are engineered to reflect a linear pathway from input to output. At the same time, the *cyclical* framing is key to assessing data sources and their implications for stakeholders in the processes where AI is brought to bear. And a *relational* perspective is required to understand the contextual motivators and impacts of AI use, and how success and risk can be most meaningfully measured.

The AI Thinking framework reflects elements of each of these framings, helping to reveal interconnections between them in practice and to find common ground within the interdisciplinary teams necessary for AI research, application, and regulation. AI Thinking thus helps to join up aspects of AI systems: illustrating how the relational networks and data cultures affect the input data used for linear computation, and interpret and use the output of that computation; and how the cyclical nature of pre-existing processes helps to shape the linear process implemented in the design of an AI technology for practical use. By understanding that these conceptualisations of AI are not in competition, but rather reflect different and equally necessary parts of the same sociotechnical systems, AI Thinking can help begin to create a cohesive model of AI use across disciplines.

### 2.3 Roots of AI Thinking: specialisation & practice

As well as drawing on a variety of disciplinary conceptualisations of AI, the AI Thinking framework explicitly builds on the practices of Statistical Thinking developed in the late nineteenth and early twentieth centuries. Similar to the way statistical methodologies matured and gave rise to wider statistical practices at the turn of the twentieth century, AI Thinking is rooted in the maturation of the AI field and its expansion from a technical specialism into a broader methodology.

Early development and use of statistical methods was primarily a matter of mathematical research, though often motivated by and entwined with the needs of states to understand changing populations (40). As statistical methods matured, however, they began to have significant value for other disciplines grappling with measuring and understanding complex phenomena, and by the mid twentieth century statistical methodologies had grown far beyond the original bounds of statistics as a specialised discipline and become essential to research and practice in psychology, medicine, and the natural sciences (41). In the later twentieth and twenty-first centuries, statistical methods have become increasingly essential to decision-making in business also (42). Whilst statistics has certainly continued to develop as a rich discipline and its own specialism, fundamentals of *statistical thinking*—considerations of sample size, recognition of random effects, uncertainty as part and parcel of research (43)—have become core competencies for contemporary research and practice across disciplines and a wide variety of applied contexts (44).

AI methods are beginning to play a similar role, providing new tools for working with the increasingly complex combinations of data and knowledge sources that characterise contemporary information systems. AI techniques, particularly machine learning, are reshaping our ability to learn from multiple information sources and combine data in complex ways. For example, in research spheres, the rapid diffusion of AI methods across research practice, from combining exascale imaging data (45) and protein folding (46) to supporting data analysis in clinical trials (47) and digital humanities (48), is moving us towards a research ecosystem where using AI methods is necessary for competitiveness in many areas of inquiry. Similar transformations are occurring in business, healthcare, policy, and other areas of information processing (15,49–51). Thus, while AI research as its own specialism will also continue to mature, fundamentals of AI Thinking are needed to help inform why, when, and how to use AI methods effectively, and to understand both their limitations and their strengths.

Like statistical thinking, AI Thinking is rooted in the professional practices and knowledge that are the foundation of training for experts in the field. Statisticians learn how to assess sample size, to measure uncertainty, to characterise populations, and so forth, and these are key building blocks with which deeper expert knowledge is built (43). Similarly, AI experts learn fundamental skills such as problem formulation (i.e., mapping a particular application to a known type of task with established methodologies), selecting and managing data, choosing models and algorithms, and interpreting the results of AI computation. But at this point, practices diverge. Expert statisticians develop deep knowledge of mechanics and principles, and may innovate with new measures and rich characterisations of complex phenomena; AI experts develop deep knowledge of mathematical and cognitive theory, and may innovate complex new model structures and learning algorithms. But for most users of these methodologies, this depth of expertise is unnecessary: the key knowledge is *operational*, knowing how to develop an appropriate process to ensure reliable evidence or synthesise information. This operational knowledge can then be combined with a deep understanding of the context in which the methods are applied, whether in specialised research, policy, decision-making, or other kinds of information processing. This bifurcation into *technical expertise* on the one hand and *operational methodology* on the other has come to characterise the dual—though actively-contested—nature of statistics in practice (52), and is a productive framing for approaching the emerging dual nature of AI.

It is important to note that AI Thinking is not the next evolution of statistical thinking, nor does it supplant statistical thinking. Indeed, as most contemporary AI relies on statistical methodologies, statistical competence is essential to AI competence, and helps AI users to cast a critical eye on the uncertainty inherent in using AI methods. AI Thinking and statistical thinking thus sit alongside one another as fundamental competencies shared across areas of working with information.

## 3. AI Thinking: An interdisciplinary, competency-based framework for AI use in practice

AI Thinking is a conceptual, competency-based model of the practices, decisions, and considerations involved in applying AI systems in context. AI Thinking is process-oriented and practice-based; competency models thus offer a natural fit for defining what AI Thinking looks like, and a practical foundation for teaching AI Thinking as a foundational information skill across disciplines and contexts (53). I outline below five core competencies of AI Thinking, representing key skills for working with AI methods as a practical information processing tool. As AI systems are typically a team effort, it is not necessary that any one person master all five of these competencies; rather, they offer a way for identifying and seeking out the skills needed in an AI team. These competencies present a starting point for further expansion into a systematic AI Thinking framework, akin to the statistical thinking model of Wild and Pfannkuch (43), and drawing on early development of professional competency frameworks for AI (49,54).

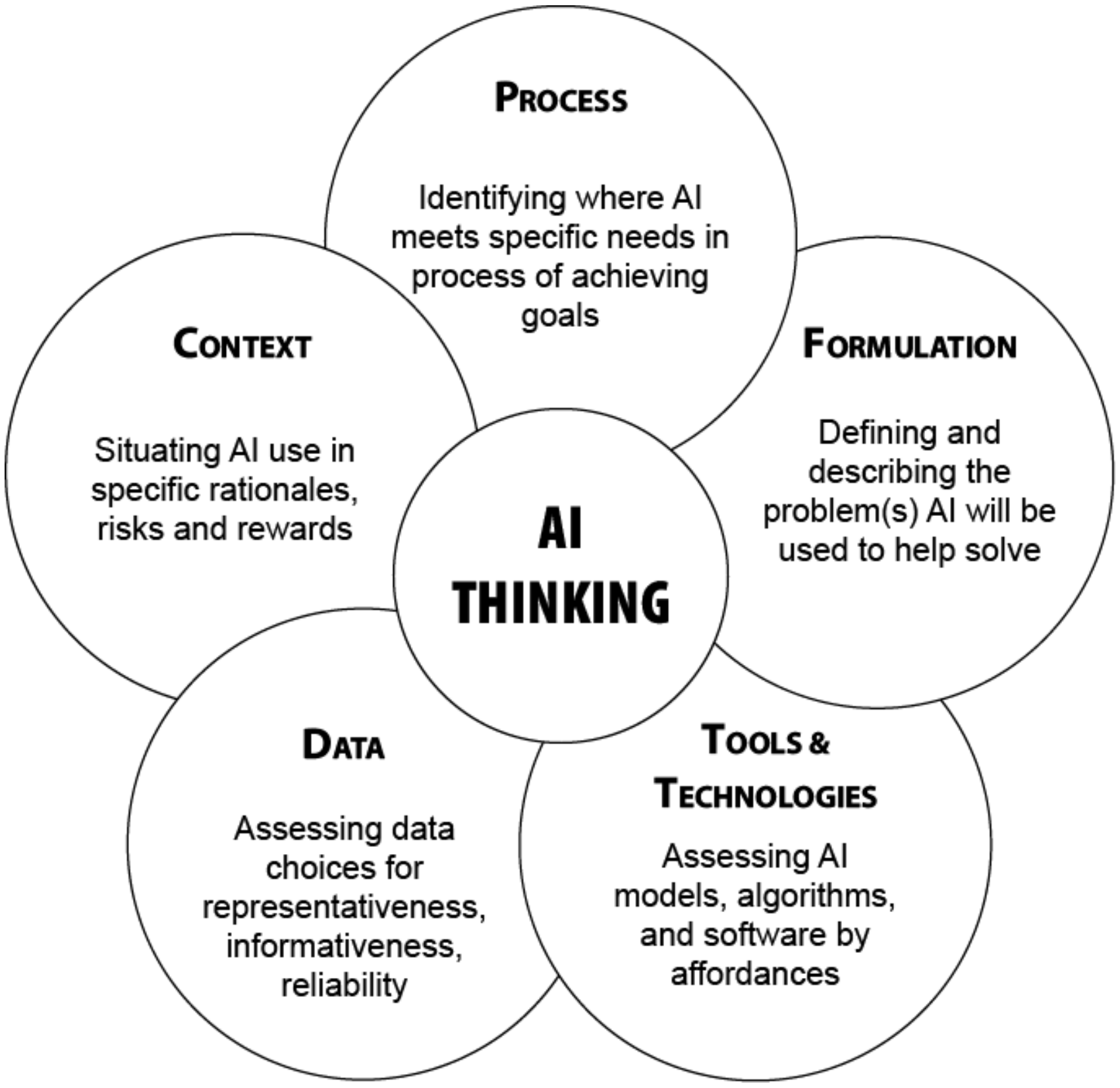

**Figure 1.** Five key competencies of AI Thinking. These competencies are arranged in a circle to show that they are interrelated and used in parallel, rather than separate concerns addressed in sequence.

The AI Thinking framework, illustrated in Figure 1, comprises five major elements of working with AI methods in practice: 1) initial alignment of AI use with ***process***-oriented needs; 2) the ***formulation*** of AI approaches to respond to those needs; 3) assessing and selecting potential AI ***tools and technologies***, to implement the chosen formulation; 4) assessing and selecting ***data*** sources to inform the chosen tools and technologies; and 5) ensuring the full lifecycle of AI system design, implementation, and management is rooted in the ***context*** of use. Figure 2

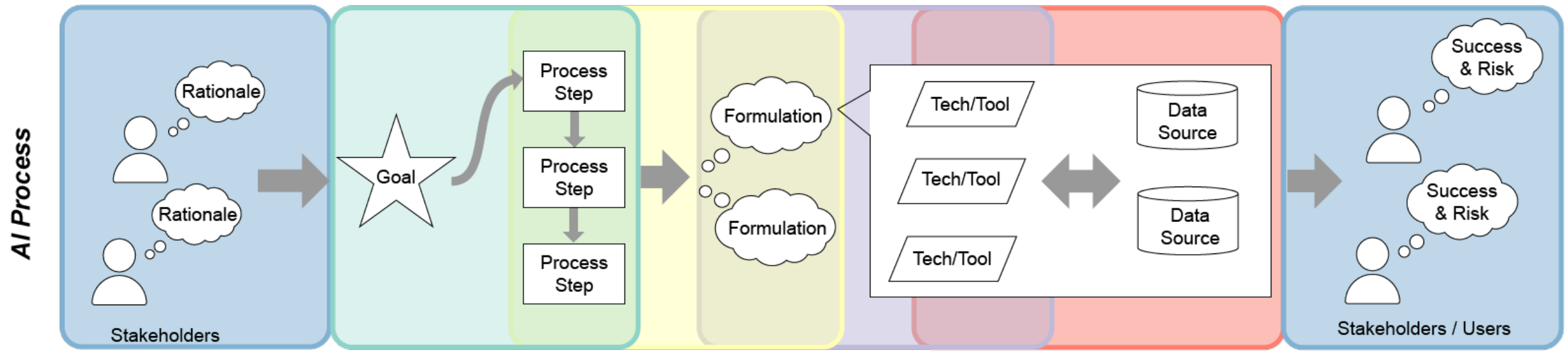

AI Thinking

| Competency | Components | |
|---|---|---|
| **Process** | • **Goal** to achieve with AI use<br>• **Steps** towatds meeting goal | • **Opportunities** for AI intervention |
| **Formulation** | • **AI task** for system to perform<br>• **Output** system will produce | • **Training signal** from which the system will learn |
| **Tools & Technologies** | • **Purpose** and AI **paradigm**<br>• **Complexity** and data needs | • **Computation demands** and benefits<br>• **Task strengths** and limitations |
| **Data** | • **Representativeness** for people and situations<br>• **Informativeness** for goal and impact | • **Reliability**, noise, and accuracy |
| **Context** | • **Stakeholders** for goal and use of AI<br>• **Rationales** of what AI should accomplish | • **Risks** of AI failure/misbehaviour<br>• **Measures of success** for broader AI benefit |

**Figure 2.** The competencies of the AI Thinking model. Each competency is aligned with one or more portions of a linearised Illustration of the process of using AI.

summarises key elements of each of these five competencies, and illustrates their interrelationships within a linearised process of AI system design.

### 3.1 Process: Motivating AI use by specific needs

Definitions of AI have changed as technologies have evolved, but AI research and innovation has been consistently characterised by a focus on *doing specific tasks*, even as AI technologies themselves have become more general-purpose (22). However, adopting AI methods in practice is often motivated by a pressure to be innovation-led and stay ahead of the technology curve, rather than as a response to specific needs for information processing that AI methods can help address (55). The greatest practical benefits from AI have historically been achieved when its use is motivated by specific needs in the service of broader goals: e.g., development of AlphaFold for identifying candidate protein structures from a vast possibility space (46); now-ubiquitous speech recognition technologies initially developed for supporting call centres (56); and automated analysis of medical imaging to support cancer diagnosis (57). In each case, a particular part of the process was identified for AI intervention, and AI use was guided by an understanding of that process.

Adopting a process-oriented motivation for AI use, and ensuring that AI applications are goal-driven rather than innovation-led, is the first key competency of AI Thinking. Taking a process-oriented approach to AI involves three distinct elements:

- AI use is *goal-driven:* developing a new AI technology or bringing an existing technology to bear is motivated by a desire to make use of information in service of a specific goal.
- AI use has a *defined scope* within the process of achieving that goal, typically in terms of specific steps or distinct operations needed. There may be multiple distinct scopes for AI use within a process, which may be approached individually or jointly.
- AI systems are designed to respond to *specific opportunities* arising within each scope for use, in terms of specific information availability and information needs at each step. Not all steps of a process will present appropriate opportunities for AI use.

This process-based approach does not preclude opportunities for innovation and discovery from using AI in a more "end-to-end" fashion to complete multiple steps or the entire process at once (discussed further below), but approaching AI use from this foundation in process clarifies opportunities and scope for intervention, provides multiple points for innovation as well as risk mitigation, and ensures that AI use is guided directly by desired goals. This approach of breaking down goals into distinct steps for AI intervention is particularly valuable when information processes need to demonstrably adhere to community norms and standards, as in research practice or business processes, or even by statutory or regulatory requirements.

Beginning from a foundation in process, AI use in practice is then informed by the next AI Thinking competency, describing and formulating the specific problem that AI use is intended to help solve.

### 3.2 Formulation: Describing the problem to solve with AI

AI is a discipline of problem-solving, and identifying how best to describe the problems to be solved has been a core component of AI development and use from its earliest days (58). *Problem formulation* involves creating formal descriptions of AI problems (typically what type of output should be produced from some type of input), and is an essential step in bridging from the complexity of real-world problems to the well-defined domain of mathematical models in AI systems. AI problem formulations have evolved from stricter mathematical design under planning paradigms of AI (59,60) to more operational definitions of tasks to perform and ways to conceptualise AI computation in current machine learning paradigms (61), but the core step of formalising the operation of an AI system remains essential for AI development and evaluation.

These shared definitions are powerful tools, making it possible to reuse and repurpose a common base of AI models and methods. For example, approaching a new, complex text labelling problem as a named entity recognition problem (in which text segments describing real-world referents are identified) enables repurposing a wide existing base of technical methods for a new purpose (62). However, as a variety of recent critical analyses have illustrated (63–65), problem formulation is highly epistemically loaded, involving normative decisions about what matters in the operation of an AI system and what that system can meaningfully say about its inputs and the world. Negotiating the tension between the benefits of shared problem formulations and the epistemic positions they represent is thus an essential part of AI use in practical contexts.

The second key competency of AI Thinking is thus understanding and negotiating the process of problem formulation for AI systems in practice. This comprises three main elements:

- The *AI task* the system is to perform, such as image classification, text generation, or speech recognition. This defines the templates by which the system should operate, and opens up opportunities for using some methods and models while closing the door on others.
- The *output* the system will produce: will it be one categorical label? Multiple numeric scores? A sequence of text? This affects how the AI system can inform its users, and what it is capable of saying about its inputs.
- If using a machine learning approach, the *training signal* from which the system will learn. This includes both what will be used to signal the desired or undesired outcome of AI processing, and the measures used to penalise errors and fit the AI model to the observed data.

Any practical information processing need where there is appropriate scope for AI intervention is likely to be open to multiple alternative problem formulations, and may in fact consist of multiple related sub-problems. This space of possible formulations and decomposition of sub-problems is illustrated by a wide range of examples of practical AI applications: Cheng et al. decomposed planning-based models for business decision-making into multiple related decision processes for greater clarity and optimisation (66); Newman-Griffis and Fosler-Lussier compare AI paradigms for categorising functional status information (67), and Isagah and Ben Dhaou describe several diverse examples of AI problem formulations in government (68).

Once a process-oriented goal for AI use has been transformed into specific problem formulations, the third AI Thinking competency addresses using the affordances offered by different AI tools and technologies to select appropriate methodologies.

### 3.3 Tools and technologies: Assessing AI affordances

As AI technologies advance at a rapid pace, keeping track of the changing landscape can seem unmanageable, and AI users often feel like new skills must be constantly relearned with each new advance (69). Comparing and contrasting available technologies for a given use of AI, and adapting and responding to new technological advancements, is thus a key skill for working with AI in practice and building sustainable AI systems that can weather the storms of technological change.

The third key competency of AI Thinking is to conceptualise AI tools and technologies in terms of the *technological affordances* they present, and how these affordances align with the intended use of AI. Technology affordances (70) extend psychological affordance theory to describe the perceived possibilities that a given technology offers to potential users, and the strengths and weaknesses it presents for different possible uses. While any specific technology may be repurposed, hacked, or extended to do a wide variety of things beyond its original intended purpose, affordances are a powerful way of characterising likely and easily perceptible opportunities a technology offers, which form a core functionality for practice. Focusing on affordances, and how they align to specific information processing needs, thus provides a clear, process-oriented strategy for comparing available AI technologies and for assessing new technologies—and the risk and reward in their adoption—when they emerge (71).

Each AI tool and technology is engineered with a particular task in mind, and each offers specific perceptible affordances for information processing. These affordances are dynamic, as new uses are discovered and reshaped, and may go far beyond the original motivating task; for example, the growth of prompt-based learning in generative AI (31), or the shift in use of AlexNet (29) and other benchmark computer vision models from image classification to broader representation of image features for many purposes (72). Assessing the affordances of AI technologies involves four key components:

- The *purpose and AI paradigm* for which a given technology was designed, including its connections to specific problem formulations as well as identifying where overlapping or complementary task definitions provide opportunities (and limitations) for interoperability and reuse.

- AI technologies have variable *complexity and data requirements* that affect their viability and effectiveness for different applications. As a general rule, more complex AI models have greater capacity to capture complex patterns and relationships, but also require larger volumes of data to work effectively. Task complexity and data availability vary widely across contexts and are essential for selecting appropriate technologies.
- AI technologies must also be assessed in terms of their *computational requirements,* ie, the hardware required to execute model calculations efficiently. While the shift to cloud computing has reduced the need to manage on-premise hardware, financial cost and environmental impact of different AI technology options are directly driven by computing requirements.
- Each technology, and each task formulation underpinning it, presents *strengths and limitations* for performing different tasks. These may be characterised by the initial research and development of new models, emerge from later, evolving uses of those models, or be considered as adaptations to an underlying framework.

Rather than assuming that newer or higher-performing technologies are necessarily better, assessing the affordances of AI technologies along these dimensions provides a balanced way to select the appropriate tool or technology for a given application. For example, in detecting and modelling rare diseases, a simpler model may be preferable to data-hungry deep learning (73), while the reverse is true for cancer detection and prognosis, where complex relationships between hundreds of variables are needed and rich multimodal data are available to learn from (74). A large language model may be an appropriate choice for processing common financial reports (75), while more bespoke analysis may be a better fit for analysing individual forms or detailed patterns of language use (76). As new technologies emerge, the same process of assessing their affordances can provide a clear pathway to determining whether they should be adopted over current approaches, and identifying what new problems they may be fit to tackle.

With a process in mind and an understanding of the affordances on offer from different AI tools and technologies to support that process, the next key competency of AI Thinking is identifying the appropriate data sources to guide AI implementation for the desired goal.

### 3.4 Data: Informing AI use with appropriate information

Data are often said to be the foundation of AI systems (16). What is less often noted is that data are also a choice, both in the process of their production and in their use in AI. The idea of "raw data," unprocessed and therefore neutral, lingers in technical discourses (23), but extensive scholarship has shown how the production of data reflects multiple choices about what counts, what to collect, and what to record about a given subject (77–79). There are also practical concerns of the compatibility of different data sources and datasets with the available tools and technologies, reflected in the well-established central role of choosing appropriate training data in machine learning research (80). In a process-oriented AI Thinking approach, the fourth key competency is thus to assess which of the available data fields, sets, and sources are most appropriate to inform the desired goals and the processes in which AI is to be used.

Assessing data sources is thus a matter of both epistemology and practicality, with each having significant power to affect bias and fairness of data and the ethical practices they reflect and act on (77). Different data choices will embed different distinctions and perspectives on the task at hand, and will work differently with available tools and technologies. While there are as many ways to assess data as there are data points, practice-based assessment of data begins with three key criteria of representativeness, informativeness, and reliability.

- *Representativeness* has many (and contested) meanings (81) but can be broadly considered in terms of data's ability to reflect the full diversity of the people or situations they are meant to convey information about.
- *Informativeness* can be distinguished from representativeness, in terms of how well data are aligned with the desired goal for AI, and how clearly they evince useful learning or patterns for that goal (82).
- *Reliability* includes noisiness as well as the likelihood of data being collected consistently and accurately in practice (83,84).

Each of these criteria may be balanced differently for different situations and goals. For example, foundation models are built on the principle of an extraordinary volume of data being sufficiently informative and representative to capture diverse general patterns despite lower reliability; in scientific analysis, data reliability may be paramount; in exploring alternative decisions a smaller set of diverse and informative data may be more valuable than data representing population norms. In addition, practical AI applications may involve multiple entry points for data with distinct balances between these criteria: eg, a diagnostic model in healthcare that begins with a foundation model and is progressively fine-tuned with smaller datasets which may be less representative of human diversity but are more informative for the diagnostic goal.

When exploring AI tools and technologies as options for intervening in a specific process, it is thus essential to assess the appropriateness and fit of different data sources for the task at hand. The final key competency of AI Thinking ties these three elements of process, tools, and data for using AI together in the specific contexts where AI is used in practice.

## 3.5 Context: Shaping AI use, benefits, and risks in situated practice

Using AI is, at its heart, about making use of computation to help us learn something, produce something, or do something in the world. As we have seen, the processes in which we bring AI to bear, the tools and technologies we choose to use, and the data sources we draw from each carry with them pieces of a broader picture in which AI use is only one element moving between different goals, actors, and perspectives. When this broader picture is ignored, or left as a separate concern from the technical design and use of AI systems, those systems inevitably miss the mark (85) and may create substantive harm (86). Recognising the contexts in which AI use is always situated, and understanding the implications of those contexts for AI's effectiveness, benefits, and risks, is thus the fifth key competency of AI Thinking.

There are, of course, many contextual aspects that bear on AI use, and analysing the sociotechnical assemblages of AI in practice is the subject of a growing research literature [cf (87–89)]. Throughout the design, implementation, and management of bringing AI methodologies into practice, a strong foundation for assessing AI contexts begins with:

- The *stakeholders* for the proposed use of AI. This includes those who set the initial goal for AI use to support, as well as those who produce the data used by AI systems and make use in turn of the outputs those systems produce.
- The *rationales* these different stakeholders have using the AI system and what it is meant to accomplish. For example, a senior leader may look to AI use for efficiency gains, while a researcher or business analyst looks to the same system to produce a specific insight from the complex data they work with.
- The *risks* posed by different points of failure or undesired behaviour in the AI system, with respect to these rationales. For example, an LLM hallucination or a spurious relationship from a machine learning model will have different impacts on an AI user, who may be concerned with accuracy in a specific instance, than on someone affected by an AI-informed decision process, for whom recourse and minimising harm are paramount.
- The *measures of success* for stakeholders in the AI system. Task-based performance metrics assess how well an AI technology performs its stated role, but may not capture the impact of that technology on someone using it as one piece among many, for example in a decision-making process.

These components of AI context may seem distant from the process of developing and implementing AI tools and technologies, but they are essential to connecting that process to the people and processes in which AI will actually be used. AI Thinking, as a model for bringing AI methodologies into practice, thus aims to connect the dots between technology, the people who use it, and the purposes it is meant to serve.

## 4. Case study of applying AI Thinking in practice

To illustrate the role of the AI Thinking framework in practice, I outline a hypothetical case study of using AI for processing information about health and well-being, an area of practice where the rich tradition of medical ethics highlights gaps in current approaches to AI ethics (90). The case study examines the use of AI methods as part of prioritisation of patients for organ transplants, and illustrates how AI application might unfold with and without AI Thinking.

### 4.1 Scenario

Prioritising potential recipients for organ donation, and ensuring successful transplantation, are highly time-sensitive and information-intensive processes. The use of rich data sources and AI systems to help tackle these challenges has been growing, and we will take as our first illustration of putting AI Thinking into practice a recent paper using machine learning-based simulations of transplant outcomes to inform decision making (91). We will imagine an extension of this methodology, implemented for prioritising potential transplant recipients in a large health system serving a demographically diverse, primarily rural population. We will consider the practices of the health informaticians building the AI system, the healthcare professionals using it to inform transplants, the patients receiving transplants, and the health system administrators assessing service outcomes and quality.

### 4.2 Without AI Thinking

Health informaticians design the AI system to tackle the motivating need directly, and rank potential recipients by outputting a single score based on an overall simulation of post-transplant health outcomes. The score is calculated based on the simulated outcome of the transplant, which is based in turn on a combination of demographic and health data, including records of past healthcare encounters. The AI technology is implemented using a single central deep neural network, which is trained based on a large volume of data sourced from the implementing health system as well as reference data from a large urban system elsewhere in the country. This is done to provide the greatest amount of data to learn from, in order to improve the model’s robustness and sensitivity to variation. The model’s performance is measured by its fidelity to past transplantation decisions.

Most healthcare professionals take the output score of the AI system and combine it with other key information, such as specific relevant lab results, distance to the patient, and size of the organ relative to the patient, to decide who to call for transplant. Some professionals mistrust the AI system and manually review health data to make recommendations, frequently arriving at different priority scores than the AI technology. Health system administrators measure success of the process using common service metrics such as 30-day mortality and rehospitalisation rate. A periodic audit after a year of using the AI system reveals that patients from urban areas are recommended for transplants at a higher rate than patients from rural areas, and that racial disparities in health outcomes are more pronounced for rural patients. These differences are consistent across professionals who use the AI system, and less pronounced for some who avoid it while being more pronounced for others. Patients and families are left with limited information about why they were not selected for transplant or why outcomes were worse than expected, with minimal transparency into the decision process or recourse to contest it.

### 4.3 With AI Thinking

Re-examining the situation with the AI Thinking framework helps to bridge gaps between understandings and practices in this scenario, and to reduce the racial and geographic disparities observed in health outcomes.

***Process**:* Prioritising transplant recipients is the overall goal, but this is implemented in professional practice by combining a number of standard key variables with an overall understanding of a patient's health status (92). Through discussion between the health informaticians and healthcare professionals, the AI technology is instead designed to target the more specific scope of assigning a transplant score to the patient's health status specifically. The other variables used in the ranking process are clearly defined by clinically-motivated rules, and not considered targets for AI intervention.

***Formulation**:* The need for transparency and accountability for health system administrators, and for insight to support clinical discretion for healthcare professionals, mean that rather than a single atomic score the task is formulated as simulating multiple component health outcomes, represented as clinically-motivated categories or continuous scores. These component outcomes are then transparently combined to an overall score by an adjustable formula. This system is trained based on the final ranking, which combines the AI-produced health score with the other, rule-based variables used by the clinicians, to model its use in practice.

***Tools and Technologies:*** Rather than a single, large deep neural network, which imposes significant data and computation requirements for the health system, the component health outcome models are implemented using a mixture of Support Vector Machines, random forests, and smaller, targeted neural network models, with significantly lower technical requirements and greater opportunity for accountability and explainability in the modelling process.

***Data**:* A review of the available data, motivated by healthcare professionals' observations, shows that a significant fraction of rural patients accessing the health system do so only through visits to the emergency department (93), and that structural racial bias emerging from the system's historical context has produced unequal delivery of services to some racial groups. To reduce the structural bias embedded into the AI system, the interdisciplinary team works together to avoid methods that assume data are race neutral (94) and minimise reliance on data that are systematically missing for rural patients.

***Context**:* Input from all stakeholders—-informaticians, healthcare professionals, health system administrators, and patients—-is incorporated throughout the application of all other competencies. System performance and outcomes are measured by monitoring mortality and rehospitalisation rates for transplant patients and explicitly measuring population disparities. The additional transparency created in the system by a process-oriented approach enables more proactive and constructive dialogue with patients and families who are affected by the system decisions.

### 4.4 Summary

In this hypothetical scenario, the AI Thinking framework provides tangible opportunities to bring together the interdisciplinary perspectives and experiences in the AI team in question. The decisions and considerations represented in the AI Thinking model clearly identify points of reflection and intervention to bridge disciplinary divides, and help to capture risks of inequitable outcomes from the system that were missed when not joining up stakeholders and disciplines. While this example might well play out differently in practice, it helps to illustrate the role of AI Thinking in working across disciplinary differences to bring AI into more thoughtful interdisciplinary use in practice.

## 5. AI Thinking in the broader AI discourse

Ongoing rapid transformations in AI have led to many different ways of framing important questions: what AI can (and should) do and why, who should use AI and how, and what is required to achieve real benefits with AI while managing its risks. This section situates AI Thinking within this broader discourse, with respect to three major reference points: AI literacy, end-to-end approaches to AI technologies, and 'AI-driven' innovation.

### 5.1 AI literacy and AI Thinking

The need for broader AI literacy is well-recognised, and strategies for developing AI literacy in students, professionals, and the public are rapidly evolving (95–97). AI Thinking, as a model for framing AI use in practice, complements current AI literacy efforts in two ways. AI literacy discussions focus primarily on formal education, whether in schools or higher education (98–100); AI Thinking provides a model to guide training and self-learning in professional and practice-based settings. AI literacy also focuses primarily on understanding how AI works and what its implications are, either from others' use or from one's own use of AI tools produced by others (97,101); AI Thinking describes the competencies required to use AI methodologies as one piece among many in the context of broader evidence-based practice. AI Thinking may be considered as an element of AI literacy, at least for practice-based audiences, but its focus on AI practice and its methodological, context-based perspective is important to distinguish from more general-purpose AI literacy efforts.

### 5.2 A process model vs "end-to-end" innovation

As fifteen years of deep learning advances have made more complex tasks more achievable for machine learning, the movement towards "end-to-end" AI approaches has continued to pick up steam. End-to-end AI treats a sequence of related tasks, each of which may be simple or difficult on its own, as a single, highly complex problem, such as in drug discovery (102), speech recognition (103), or even autonomous driving (104); this allows (it is argued) for greater innovation and discovery of unanticipated relationships and approaches to solving difficult problems. AI Thinking, as a process-oriented approach involving decomposing goals into multiple specific points for AI intervention, seems at first to run counter to the end-to-end model. However, a process-based orientation does not necessitate a process-based intervention: with appropriate AI tools and technologies and supporting data sources available, an end-to-end approach may well be supported under the AI Thinking model. What AI Thinking provides is a clearer framework for identifying whether an end-to-end or a step-by-step approach is most appropriate to the specific goal, tools, data, and context at hand. Process-based AI Thinking can also help to identify clearer points for evaluation, control, and risk mitigation in complex AI systems, helping to tackle some of the longstanding problems in end-to-end approaches (105).

### 5.3 AI Thinking or AI doing? Being AI-informed rather than AI-driven

The idea of "AI-driven" innovation has gained common currency in recent years, in diverse settings including policy (106), science (107), and business (51). While the meaning of the term varies in practice, an "AI-driven" framing creates a persistent narrative of innovation and discovery being *led* directly by the use of AI technologies, as opposed to AI *supporting* the process as one tool among many. The framing of "AI Thinking," rather than a more action-oriented "AI doing," emphasises the role of AI as a support tool, and reflects the purpose of the model to describe the design considerations and decisions made in setting up the use of AI, before that informed use then leads to specific action. AI Thinking is thus more oriented towards *AI-informed* design and use, in which AI functions as part of a larger sociotechnical system (108).

## 6. Conclusion

AI methodologies are transforming the way we work with information. A wide range of disciplines study, use, and develop AI and its transformations, but the different perspectives on AI and its use adopted by these disciplines are often in conflict. New approaches are needed to effectively combine the practical insights and analytic methods offered by different disciplines, and to build more holistic interdisciplinary practices informing AI applications in context.

In this article, I have proposed a novel conceptual framework called *AI Thinking*, which models key decisions and considerations to help bridge disciplinary perspectives on AI in practice. Just as statistical thinking transformed our use and understanding of growing information in the nineteenth and twentieth centuries (109), so AI Thinking can help transform information processing in the twenty-first century. I have presented a competency-based model of AI Thinking, representing the key practices involved in applying AI in context. These five competencies—orienting AI use in process, formulating AI implementations, assessing the affordances of AI tools and technologies, selecting appropriate data sources, and situating AI use in context—provide a starting point for rethinking AI skills and training to facilitate more interdisciplinary application of AI, as well as bridging disciplinary perspectives on how AI systems can be better designed, assessed, and managed.

As AI Thinking is brought more systematically into practice, and AI use continues to become more interdisciplinary in nature, the initial competencies presented here will be further tested and expanded. Wider empirical study of AI practices and needs will be essential to further develop the AI Thinking model and to keep it aligned with evolving AI practice. As the AI discourse continues to grow and change, the relationships between AI Thinking, AI literacy, and other key framings in AI innovation and practice outlined here will further develop, enabling new connections and comparisons to inform richer conceptualisations of AI in practice. Strengthening dialogue and collaboration on AI practices across disciplines is vital to achieving practical benefits from AI use while reducing and managing its risks. By bridging conceptual and disciplinary divides in AI research and practice, the AI Thinking framework aims to help shape a more purpose-driven, effective, and manageable future of AI in practice.

## Acknowledgments

I am deeply grateful to Anna Newman-Griffis and Susan Oman for invaluable discussions of language and framing and provocations to the framework which have helped shape the arguments presented here, and to Xin Zhao, Andrew M. Cox, Michael Thewall, Stephen Pinfield, and Ben Steyn for discussions and feedback on early drafts of this article. Many thanks to the organisers and attendees of WOEPSR 2023, and to the teams, Steering Groups, and partners of the GRAIL and FRAIM projects for wider discussions on key ideas.

The author declares that this work was completed in the absence of competing interests.

## Funding

The research in this article was supported in part by the Framing Responsible AI Implementation and Management (FRAIM) project, funded by the Bridging Responsible AI Divides (BRAID) programme of the UKRI Arts and Humanities Research Council (AHRC; award AH/Z505596/1). It was also supported as part of the GRAIL project of the Research on Research Institute (RoRI). RoRI's second phase (2023–2027) is funded by an international consortium of partners, including: Australian Research Council (ARC); Canadian Institutes of Health Research (CIHR); Digital Science; Dutch Research Council (NWO); Gordon and Betty Moore Foundation [Grant number GBMF12312; DOI 10.37807/GBMF12312]; King Baudouin Foundation; La Caixa Foundation; Leiden University; Luxembourg National Research Fund (FNR); Michael Smith Health Research BC; National Research Foundation of South Africa; Novo Nordisk Foundation [Grant number NNF23SA0083996]; Research England (part of UK Research and Innovation); Social Sciences and Humanities Research Council of Canada (SSHRC); Swiss National Science Foundation (SNSF); University College London (UCL); Volkswagen Foundation; and Wellcome Trust [Grant number 228086/Z/23/Z]. Sincere thanks to all our partners for their engagement and support.

Responsibility for the content of RoRI-supported outputs lies with the authors and RoRI CIC. The views expressed in this article do not necessarily reflect those of RoRI's partners, the BRAID programme, or AHRC. RoRI is committed to open research as an important enabler of our mission, as set out in our Open Research Policy. Any errors or omissions remain our own.